\documentstyle[11pt,aaspp4,flushrt]{article} 

\begin{document}
\title{Stability in Force-Free Electrodynamics} 
\author{Andrei Gruzinov}
\affil{Institute for Advanced Study, School of Natural Sciences, Princeton, NJ 08540}

\begin{abstract}
We show that flows of pure electromagnetic energy are subject to instabilities familiar from hydrodynamics and plasma physics---Kelvin-Helmholtz and screw instabilities. In the framework of force-free electrodynamics, it is found that electric-like  tangential discontinuities are Kelvin-Helmholtz-unstable. All non-trivial cylindrically symmetrical magnetic configurations are screw-unstable. The Poynting jet of the Goldreich-Julian pulsar is screw-unstable if the current density in the polar cap region exceeds the charge density times the speed of light. This may be the process that determines the pulsar luminosity.

\end{abstract}
\keywords{instabilities $-$ magnetic fields}

\section{Introduction}
It has been suggested that the luminosity of pulsars, Kerr black holes in external magnetic fields, relativistic accretion disks, and gamma-ray bursts comes originally in the form of a Poynting jet. We show that Poynting jets are subject to Kelvin-Helmholtz and screw instabilities. Generically, Poynting jets should lose their stability as they propagate away from the sources because the guiding magnetic field decreases more quickly than the toroidal magnetic field. Both the guiding magnetic field $B$ and the Poynting flux density $S$ decrease as $1/A$, where $A$ is the cross section of the Poynting jet, and we assume that $A$ grows as the jet propagates away from the source. Since the toroidal magnetic  field and the radial (in local cylindrical coordinates) electric field scale as $S^{1/2}$, these fields decrease slower than the guiding field $B$. The stabilizing effect of the guiding field decreases, causing the jet to lose stability. Instabilities probably lead to a turbulent cascade of energy to smaller scales. The ultimate energy sink may be particle acceleration and/or pair production. 

We investigate the instabilities in the framework of force-free electrodynamics (FFE). FFE has been applied to pulsars, black holes, accretion disks, and gamma-ray bursts (Goldreich \& Julian 1969, Scharlemann \& Wagoner 1973, Blandford 1976, Blandford \& Znajek 1977, Michel 1973, Thompson \& Blaes 1998, Meszaros \& Rees 1997). By definition, FFE applies when the energy-momentum tensor is dominated by the electromagnetic fields. Particles carry charges and currents but have zero inertia. We present in \S 2 different formulations of FFE that are used in the stability calculations. We show in \S 3 that electric-like tangential discontinuities are Kelvin-Helmholtz unstable. We prove in \S 4 that all non-trivial cylindrically symmetrical magnetic configurations are screw unstable. We give in \S 5.1 the eigenmode equation  for a cylindrical Poynting jet. As an illustration of applications of this equation, we show in \S 5.2 that the Poynting jet of the Goldreich-Julian pulsar is screw unstable if the Poynting luminosity exceeds the classical ``magneto-dipole'' value. 

\section{Force-free electrodynamics (FFE)}
FFE is applicable if electromagnetic fields are strong enough to produce pairs and baryon contamination is prevented by strong gravitational fields (Blandford \& Znajek 1977). Pulsars, Kerr black holes in external magnetic fields, relativistic accretion disks, and gamma-ray bursts are the astrophysical objects whose luminosity might come originally in a pure electromagnetic form describable by FFE. 

FFE is classical electrodynamics supplemented by the force-free condition:
\begin{equation}
\partial _t{\bf B}=-\nabla \times {\bf E},
\end{equation}
\begin{equation}
\partial _t{\bf E}=\nabla \times {\bf B}-{\bf j},
\end{equation}
\begin{equation}
\rho {\bf E}+{\bf j}\times {\bf B}=0.
\end{equation}
$\nabla \cdot {\bf B}=0$ is the initial condition. The speed of light is $c=1$; $\rho =\nabla \cdot {\bf E}$ and ${\bf j}$ are the charge and current densities multiplied by $4\pi$. The electric field is everywhere perpendicular to the magnetic field, ${\bf E}\cdot {\bf B}=0$ \footnote{If $\rho \not= 0$, ${\bf E}\cdot {\bf B}=0$ follows from (3), if $\rho =0$, this condition is an independent basic equation of FFE.}. The electric field component parallel to the magnetic field should vanish because the charges are freely available in FFE. It is also assumed that the electric field is everywhere weaker than the magnetic field, $E^2<B^2$. Then equation (3) means that it is always possible to find a local reference frame where the field is a pure magnetic field, and the current is flowing along this field. FFE is Lorentz invariant. 

Equation (3) can be written in the form of the Ohm's law. The current perpendicular to the local magnetic field can be calculated from equation (3). The parallel current is determined from the condition that electric and magnetic fields remain perpendicular during the evolution described by the Maxwell equations (1), (2).  We thus obtain the following non-linear Ohm's law
\begin{equation}
{\bf j}={({\bf B}\cdot \nabla \times {\bf B}-{\bf E}\cdot \nabla \times {\bf E}){\bf B}+(\nabla \cdot {\bf E}){\bf E}\times {\bf B} \over B^2}.
\end{equation}
Equations (1), (2), (4) form an evolutionary system (initial condition ${\bf E}\cdot {\bf B}=0$ is assumed). It therefore makes sense to study stability of equilibrium electromagnetic fields in FFE. One can also study linear waves and their nonlinear interactions in the framework of FFE (Thompson \& Blaes 1998). 

It is convenient to introduce a formulation of FFE similar to magnetohydrodynamics (MHD); then we can use the familiar techniques of MHD to test stability of magnetic configurations. To this end, define a field ${\bf v}={\bf E}\times {\bf B}/B^2$, which is similar to velocity in MHD. Then ${\bf E}=-{\bf v}\times {\bf B}$ and equation (1) becomes the ``frozen-in'' law
\begin{equation}
\partial _t{\bf B}=\nabla \times ({\bf v}\times {\bf B}).
\end{equation}
From ${\bf v}={\bf E}\times {\bf B}/B^2$, and from equations (1)-(3), one obtains the momentum equation
\begin{equation}
\partial _t(B^2{\bf v})=\nabla \times {\bf B}\times {\bf B}+\nabla \times {\bf E}\times {\bf E}+(\nabla \cdot {\bf E}){\bf E},
\end{equation}
where ${\bf E}=-{\bf v}\times {\bf B}$. Equations (5), (6) are the usual MHD equations except that the density is equal $B^2$ and there are order $v^2$ corrections in the momentum equation.

\section{Tangential discontinuities}

The planar Poynting jet ${\bf E}_0=(V(x),0,0)$, ${\bf B}_0=(0,U(x),B(x))$ is a stationary solution of FFE if
\begin{equation}
B^2+U^2-V^2={\rm const}.
\end{equation}
The eigenmodes of this jet are $\propto \exp (-i\omega t+ik_yy+ik_zz)$. As shown in Appendix A, the eigenmode equation for the x-displacement $\xi$, defined by $\delta B_x=i(Uk_y+Bk_z)\xi$, is 
\begin{equation}
(F\xi ')'+(\omega ^2-k_y^2-k_z^2)F\xi =0.
\end{equation}
Here the prime denotes the x-derivative, and $F=(\omega B+k_yV)^2-(k_zB+k_yU)^2+(k_zV-\omega U)^2$.

To illustrate applications of the eigenmode equation (8), consider a tangential discontinuity ${\bf E}_0=(V,0,0)$, ${\bf B}_0=(0,U,B)$ at $x<0$, and ${\bf E}_0=0$, ${\bf B}_0=(0,0,B_0)$ at $x>0$. As follows from equation (7), $B_0^2=B^2+U^2-V^2$. The eigenmode equation (8) is solved by $\xi=\exp (-\kappa |x|)$, $\kappa ^2=k_y^2+k_z^2-\omega ^2$. From the continuity of $F\xi '$, $F(-0)+F(+0)=0$, which gives the dispersion law
\begin{equation}
(\omega B+k_yV)^2-(k_zB+k_yU)^2+(k_zV-\omega U)^2+(\omega ^2-k_z^2)B_0^2=0.
\end{equation}
 
It follows from the dispersion law (9) that magnetic-like discontinuities, $|U|>|V|$, are stable and electric-like discontinuities , $|V|>|U|$, are unstable \footnote{The simplest way to see this is to use a Lorentz boost along $z$ to remove either $U$ or $V$.}.

\section{Magnetic configurations}

A force-free magnetic field ($\nabla \times {\bf B}\times {\bf B}=0$) with a zero electric field is a stationary FFE solution. We will study the stability of the force-free magnetic configurations using the MHD formulation of FFE given by equations (5) and (6). Linear perturbations, denoted ${\bf b}$ and ${\bf v}$, $\propto \exp (-i\omega t)$, satisfy
\begin{equation}
-i\omega {\bf b}=\nabla \times ({\bf v}\times {\bf B}),
\end{equation}
\begin{equation}
-i\omega B^2{\bf v}=\nabla \times {\bf B}\times {\bf b}+\nabla \times {\bf b}\times {\bf B}.
\end{equation}
Define the displacement \mbox{\boldmath $\xi$} by ${\bf v}=-i\omega \mbox{\boldmath $\xi$}$. From (10),  ${\bf b}=\nabla \times (\mbox{\boldmath $\xi$}\times {\bf B})$. Now equation (11) can be written as
\begin{equation}
-\omega ^2B^2\mbox{\boldmath $\xi$}=\nabla \times {\bf B}\times \nabla \times (\mbox{\boldmath $\xi$}\times {\bf B})+\nabla \times \nabla \times (\mbox{\boldmath $\xi$}\times {\bf B})\times {\bf B}\equiv -\hat{\bf K} \mbox{\boldmath $\xi$}.
\end{equation}

Since the operator $\hat{\bf K}$ is self-adjoint (Kadomtsev 1966 and references therein), the frequency is given by the variational principle
\begin{equation}
\omega ^2={\rm min}{ \int d^3r\mbox{\boldmath $\xi$}\hat{\bf K}\mbox{\boldmath $\xi$} \over \int d^3rB^2\mbox{\boldmath $\xi$}^2 }.
\end{equation}
The equilibrium field ${\bf B}$ is unstable if the potential energy
\begin{equation}
W\equiv \int d^3r\mbox{\boldmath $\xi$}\hat{\bf K}\mbox{\boldmath $\xi$},
\end{equation}
is negative for some displacements \mbox{\boldmath $\xi$}. 

Consider the case of cylindrical symmetry, coordinates $(r,\theta ,z)$. The equilibrium field ${\bf B}=(0,U(r),B(r))$ should satisfy $BB'+Ur^{-1}(rU)'=0$, where the prime denotes the r-derivative. For an eigenmode $\propto \exp (im\theta +ikz)$, the potential energy reduces to (Kadomtsev 1966)
\begin{equation}
W\equiv \int dr(f\xi '^2+g\xi ^2),
\end{equation}
where $\xi $ is the radial component of the displacement. The other two components of the displacement vector, $\xi _{\theta }$ and $\xi _z$, were chosen to minimize the energy for a given $\xi $. The functions of radius $f$ and $g$ are given by
\begin{equation}
f=r{(krB+mU)^2\over k^2r^2+m^2},
\end{equation}
\begin{equation}
g=r^{-1}\left( {k^2r^2+m^2-1\over k^2r^2+m^2}(krB+mU)^2+{2k^2r^2\over (k^2r^2+m^2)^2}(k^2r^2B^2-m^2U^2)\right).
\end{equation}

We now show that magnetic configurations with a non-zero toroidal field $U$ are screw unstable. Screw means, e. g., that $m=1$, but not $m=-1$, is unstable for a given sign of $k$. Kadomtsev (1966) gives a clear discussion of the screw instability in plasmas, and shows that the screw mode is the most dangerous mode (if the plasma is unstable, it is screw unstable).    

Assume that $U(0)=0$, $U'(0)>0$, $U\rightarrow 0$ for $r\rightarrow \infty$, and $U$ is positive in between. Assume that $B$ is everywhere positive. Let $k$ be positive and small. Take $m=-1$. Let $r_0$ be the first zero of $krB-U$. Take a trial function $\xi=1$ for $r<r_0$, and zero at $r>r_0$. We can chose this generalized function $\xi $ in such a way that the first term in the energy integral (15) vanishes. The second term is an integral of $g$ from $0$ to $r_0$. For small $k$, $g=k^2r(krB-U)(3krB+U)$ is negative. 

\section{Poynting jets}
Poynting jets become observable only after the instabilities convert the Poynting flux into kinetic energy of charged particles. The FFE instabilities may also limit the Poynting luminosity of the sources. Thus, the stability of Poynting jets in FFE is of interest for astrophysical applications. Unlike the purely magnetic case, the eigenvalue problem is not self-adjoint. We first give the eigenmode equation for cylindrical Poynting jets. We then analyze stability of a slowly rotating Goldreich-Julian pulsar. We will show that the Poynting jet is screw unstable if the current density exceeds the charge density times the speed of light \footnote{Since charges of either sign are available, the current density can, in principle, exceeds the charge density times the speed of light.}. 

\subsection{The eigenmode equation}
Cylindrical Poynting jet ${\bf E}_0=(V(r),0,0)$, ${\bf B}_0=(0,U(r),B(r))$, in cylindrical coordinates $(r,\theta ,z)$, is a stationary solution of FFE if
\begin{equation}
BB'+Ur^{-1}(rU)'-Vr^{-1}(rV)'=0,
\end{equation}
the prime denotes the r-derivative. The eigenmodes are $\propto \exp (-i\omega t+im\theta +ikz)$. Proceeding along the same lines as in Appendix A, we obtain the eigenmode equation for the r-displacement $\xi$, defined by $\delta B_r=i(mU/r+Bk)\xi$,
\begin{equation}
(f\xi ')'=g\xi .
\end{equation}
Here 
\begin{equation}
f=-rF/\kappa ^2,
\end{equation}
\begin{equation}
g=r^{-1}E/\kappa ^2,
\end{equation}
\begin{equation}
\kappa ^2=k^2+m^2r^{-2}-\omega ^2,
\end{equation}
\begin{equation}
F=c_1^2-c_2^2+c_3^2,
\end{equation}
\begin{equation}
E=(1-\kappa ^2r^2)F-2{k^2-\omega ^2\over \kappa ^2}G-4H-rH',
\end{equation}
\begin{equation}
G=2B(\omega c_1-kc_2)-F,
\end{equation}
\begin{equation}
H=c_3^2,
\end{equation}
\begin{equation}
c_1=\omega B+mV/r,
\end{equation}
\begin{equation}
c_2=kB+mU/r,
\end{equation}
\begin{equation}
c_3=kV-\omega U.
\end{equation}

The eigenmode equation (19) reduces to (8) in the planar limit, $r\rightarrow \infty$, $m\rightarrow \infty$, $m/r={\rm const}=k_y$. In the purely magnetic case, $V=0$, for $\omega =0$, the functions $f$ and $g$ reduce to (16), (17). The eigenmode equation is invariant under the Lorentz boosts in the z-direction - since $(\omega , k)$, $(U,V)$, and $(c_1, c_2)$ are vectors, one can show that $f$ and $g$ are scalars. 

\subsection{Stability of the Goldreich-Julian pulsar model}
To illustrate how equation (19) can be applied, we will derive the stability criterion for a slowly rotating Goldreich-Julian (1969) pulsar. The pulsar is called slow if $\Omega R\ll c$, where $\Omega$ is the neutron star (NS) rotation frequency and $R$ is the NS radius, $c=1$ is the speed of light. The pulsar luminosity is carried by a narrow Poynting jet, radius $a\sim (\Omega R/c)^{1/2}R$, emanating from the polar cap region. Since $a\ll R$, the jet is approximately cylindrical. In the vicinity of the axis, the electrostatic potential is given by $V=\Omega Br$, $B$ is the NS magnetic field. The toroidal magnetic field can be written in the form $U=\Omega _U (r)Br$. We will assume that $\Omega _U (r)$ decreases from some positive value $\Omega _U$ at $r=0$ to $0$ at $r>a$. The value of $\Omega_U$ is not known, but it was suggested that $\Omega _U \sim \Omega$. 

The following analytical derivation of the sufficient condition for screw instability has been confirmed by numerical simulations of the eigenmode equation. Take $k=(1-\epsilon )\Omega _U$, where $\epsilon$ is a small positive number. If $\Omega =0$, the screw mode ($m=-1$) is unstable and has a small growth rate. What happens when we ``turn on'' $\Omega$? Since $f/g\sim 1/k^2\gg a^2$, the eigenmode is approximately a step-function, $\xi =1$ at $r<r_0$, $\xi =0$ at $r>r_0$, where $r_0$ is the first zero of $c_2$, and the real part of the eigenfrequency is ${\rm Re}~\omega =-\Omega$, so that $|c_1|\ll |c_2|$. The instability criterion is similar to the purely magnetic case of \S 4, 
\begin{equation}
\int _0^{r_0}dr g<0.
\end{equation}
In the leading order in $\Omega R/c$,
\begin{equation}
g=(k^2-\Omega ^2)(kB-U/r)^2+2(k^2-\Omega ^2)(k^2B^2-U^2/r^2)-3\Omega ^2(kB-U/r)^2.
\end{equation} 
The second term in $g$ is the largest in absolute value. This term is negative if $\Omega < \Omega _U$. Therefore the jet is unstable so long as  $\Omega < \Omega _U$. It follows that the luminosity of a stable Goldreich-Julian pulsar does not exceed $L\sim a^2cUV\sim (\Omega R/c)^4cB^2R^2$ - the ``magneto-dipole'' value.

\acknowledgements I thank John Bahcall and Roger Blandford for discussions. This work was supported by NSF PHY-9513835. 

\begin{appendix}

\section{Planar symmetry}
We derive in this Appendix the eigenmode equation (8). Consider a planar Poynting jet ${\bf E}_0=(V(x),0,0)$, ${\bf B}_0=(0,U(x),B(x))$, $BB'+UU'-VV'=0$, the prime denotes the x-derivative. The eigenmodes are $\propto \exp (-i\omega t+ik_yy+ik_zz)$. The FFE equations (1)-(3) for the perturbed fields and currents (denoted $e$, $b$, $j$) are 
\begin{equation}
\omega b_x=k_ye_z-k_ze_y,
\end{equation}
\begin{equation}
\omega b_y=k_ze_x+ie_z',
\end{equation}
\begin{equation}
\omega b_z=-k_ye_x-ie_y',
\end{equation}
\begin{equation}
\omega e_x=-k_yb_z+k_zb_y-ij_x,
\end{equation}
\begin{equation}
\omega e_y=-k_zb_x-ib_z'-ij_y,
\end{equation}
\begin{equation}
\omega e_z=k_yb_x+ib_y'-ij_z,
\end{equation}
\begin{equation}
V'e_x+V(e_x'+ik_ye_y+ik_ze_z)-B'b_z-U'b_y+Bj_y-Uj_z=0,
\end{equation}
\begin{equation}
V'e_y+U'b_x-Bj_x=0,
\end{equation}
\begin{equation}
V'e_z+B'b_x+Uj_x=0.
\end{equation}
From (A8), (A9), 
\begin{equation}
Vb_x+Ue_y+Be_z=0.
\end{equation}
From (A4), (A8),
\begin{equation}
V'e_y+U'b_x=iB(\omega e_x+k_yb_z-k_zb_y).
\end{equation}
From (A5), (A6), (A7)
\begin{equation}
(Bb_z+Ub_y-Ve_x)'=i(c_1e_y+c_2b_x+c_3e_z),
\end{equation}
where we denoted $c_1=\omega B+k_yV$, $c_2=k_zB+k_yU$, $c_3=k_zV-\omega U$. For further derivations the following relationships are useful: $c_1U-c_2V+c_3B=0$, and $c_1k_z-c_2\omega -c_3k_y=0$. 

We solve (A1), (A10) by introducing the x-displacement $\xi$ (called the x-displacement because according to (A13) $b_x$ is obtained from $U$ and $B$ by an x-displacement $\xi$):
\begin{equation}
b_x=ic_2\xi ,~~~~~e_y=-ic_1\xi ,~~~~~e_z=-ic_3\xi .
\end{equation}
Define
\begin{equation}
\eta =Bb_z+Ub_y-Ve_x.
\end{equation}
Equation (A12) takes the form
\begin{equation}
\eta '=(c_1^2-c_2^2+c_3^2)\xi .
\end{equation}
From (A2), (A3), (A13), (A14), (A11)
\begin{equation}
B(c_1\xi )'=-(\omega V+k_yB)e_x+\omega Ub_y-\omega \eta ,
\end{equation}
\begin{equation}
(c_3\xi )'=-k_ze_x+\omega b_y,
\end{equation}
\begin{equation}
(c_2U'-c_1V')\xi =c_1e_x-c_2b_y+k_y\eta .
\end{equation}
Excluding $e_x$ and $b_y$ from (A16)-(A18), we obtain 
\begin{equation}
(c_1^2-c_2^2+c_3^2)\xi '=(k_y^2+k_z^2-\omega ^2)\eta .
\end{equation}
From (A15), (A19),
\begin{equation}
((c_1^2-c_2^2+c_3^2)\xi ')'=(k_y^2+k_z^2-\omega ^2)(c_1^2-c_2^2+c_3^2)\xi .
\end{equation}

\end{appendix}

\end{document}